\documentclass[aps,superscriptaddress,twocolumn]{revtex4-2}

\usepackage{amssymb,amsmath,amsfonts,mathrsfs,bm, bbm, graphicx, epsfig, epstopdf,soul,physics,cancel}
\usepackage[version=3]{mhchem}
\usepackage{braket}
\usepackage{hyperref}
\usepackage{color}
\usepackage{csquotes}
\usepackage{float}

\begin{document}
\title{Proposal for non-cryogenic quantum repeaters with hot hybrid alkali-noble gases}
\author{Jia-Wei Ji}
\email{quantum.jiawei.ji@gmail.com}
\affiliation{Institute for Quantum Science and Technology, and Department of Physics \& Astronomy, University of Calgary, 2500 University Drive NW, Calgary, Alberta T2N 1N4, Canada}

\author{Faezeh Kimiaee Asadi}
\affiliation{Institute for Quantum Science and Technology, and Department of Physics \& Astronomy, University of Calgary, 2500 University Drive NW, Calgary, Alberta T2N 1N4, Canada}

\author{Khabat Heshami}
\affiliation{National Research Council of Canada, 100 Sussex Drive, Ottawa, Ontario K1A 0R6, Canada}

\affiliation{Department of Physics, University of Ottawa, Advanced Research Complex, 25 Templeton Street, Ottawa, Ontario K1N 6N5, Canada}

\author{Christoph Simon}
\email{christoph.simon@gmail.com}
\affiliation{Institute for Quantum Science and Technology, and Department of Physics \& Astronomy, University of Calgary, 2500 University Drive NW, Calgary, Alberta T2N 1N4, Canada}

\begin{abstract}
We propose a quantum repeater architecture that can operate without cryogenics. Each node in our architecture builds on a cell of hot alkali atoms and noble-gas spins which offer a storage time as long as a few hours. Such a cell of hybrid gases is placed in a ring cavity, which allows us to suppress the detrimental four-wave mixing (FWM) noise in the system. We investigate the protocol based on a single-photon source made of an ensemble of the same hot alkali atoms. A single photon emitted from the source is either stored in the memory or transmitted to the central station to be detected. We quantify the fidelity and success probability of generating entanglement between two remote ensembles of noble-gas spins by taking into account finite memory efficiency, channel loss, and dark counts in detectors. We describe how the entanglement can be extended to long distances via entanglement swapping operations by retrieving the stored signal. Moreover, we quantify the performance of this proposed repeater architecture in terms of repeater rates and overall entanglement fidelities and compare it to another recently proposed non-cryogenic quantum repeater architecture based on nitrogen-vacancy (NV) centers and optomechanical spin-photon interfaces. As the system requires a relatively simple setup, it is much easier to perform multiplexing, which enables achieving rates comparable to the rates of repeaters with NV centers and optomechanics, while the overall entanglement fidelities of the present scheme are higher than the fidelities of the previous scheme. Our work shows that a scalable long-distance quantum network made of hot hybrid atomic gases is within reach of current technological capabilities.
\end{abstract}

\maketitle

\section{Introduction}\label{sec:introduction}
The realization of global quantum networks would bring many fascinating applications to the world, which include secure communication~\cite{RevModPhys.74.145}, blind quantum computing~\cite{Barz303}, private database queries~\cite{PhysRevA.83.022301}, and eventually, a quantum internet that connects quantum computers and other quantum information processing devices \cite{Kimble2008,Simon2017,Wehnereaam9288}. In such a quantum network, photons are used as information carriers for establishing long-distance connections, but they are adversely affected by transmission loss, which significantly limits the distance of connecting remote locations. Unlike its classical counterparts, photon loss cannot be compensated by amplification as unknown quantum states cannot be perfectly cloned according to the no-cloning theorem \cite{Wootters1982}. Therefore, quantum repeaters have been proposed to solve this issue but this requires stationary quantum memories for storing and processing the quantum information \cite{RevModPhys.83.33,Muralidharan2016,Simon2017}. Currently, a vast majority of approaches to quantum networks need either vacuum equipment and optical trapping or cryogenic cooling~\cite{Duan2001,Kumar_2019,KimiaeeAsadi2018quantumrepeaters,PhysRevLett.123.063601,Humphreys2018,RevModPhys.83.33,Delteil2016,PhysRevLett.119.010503,Asadi_2020}, which makes scaling up such architectures very difficult. However, there have been some efforts in proposing quantum networks that operate at room temperature based on solid-state systems \cite{Ghobadi2019,Ji2022proposalroom} but they require complex setups and have high demands in designing the hardware for realizing the spin-photon interface. On the other hand, hot alkali vapors have been actively investigated as quantum memories for the application of quantum networks \cite{Borregaard2016,li2021heralding,zugenmaier2018long,dou2018broadband,dideriksen2021room}, and as they require relatively simple setups, it is easier to scale, which even offers a great potential for being deployed in space \cite{gundougan2021topical}.

In spite of the appealing features of hot alkali vapor, there are a few challenges in the system. The main roadblock towards using this system for quantum networks is four-wave mixing (FWM) noise as it is quite significant and ubiquitous in $\Lambda$-type hot atomic ensembles, posing serious challenges to the single-photon level applications  \cite{phillips2008optimal,lauk2013fidelity}. Proposed solutions to this issue include blocking FWM channels by polarization selection rules \cite{zhang2014suppression}, Raman absorption-enabled suppression in a mixed hot vapor \cite{romanov2016suppression}, cavity engineering \cite{nunn2017theory}, and by means of coherent destructive interference of FWM \cite{thomas2019raman}. The advantage of using a cavity to suppress FWM noise compared to other solutions is that it offers enhanced light storage and retrieval efficiency while only introducing a cavity. It has been experimentally verified, reporting a noise floor of around $1.5\times 10^{-2}$ photons per pulse in a Raman-type hot vapor memory \cite{saunders2016cavity}. Another significant challenge in the $\Lambda$-type hot atomic ensembles is short storage time in the collective spin state, which is mainly affected by the atomic collisions between the hot vapor and the buffer gas and the collisions in the hot vapor itself. Due to this detrimental effect, the storage time in hot ensembles is limited to a microsecond \cite{hammerer2010quantum}, thus restricting its application in quantum networks. However, there has been some work towards reducing this detrimental effect either by the motional averaging method \cite{Borregaard2016,dideriksen2021room} or by using a decoherence-free subspace of spin states \cite{katz2018light} with the spin coherence time extended to a second but even second-long coherence time may not be sufficient for long-distance quantum networks \cite{RevModPhys.83.33}.

Rare isotopes of noble gas have non-zero nuclear spins, which are isolated from the environment by electronic shells. Thus, they maintain hours-long coherence time even at room temperature~\cite{gentile2017optically}. They can be accessed either via the collisions with metastable helium atoms or via the collisions with alkali atoms~\cite{gentile2017optically}. A quantum interface between noble-gas spins and alkali atoms has been proposed based on weak spin-exchange collisions~\cite{PRXQuantum.3.010305}. Using this interface, the storage time can be significantly enhanced, which has been experimentally demonstrated with the coherence time of a minute \cite{katz2021coupling} and an hour \cite{shaham2022}.

In this work, we propose a quantum repeater architecture without cryogenics, which is based on hot alkali vapor and noble-gas nuclear spins. In our proposal, we adopt the cavity engineering method to suppress FWM noise when the input gets stored as a collective spin excitation in hot vapor via the off-resonant Raman protocol \cite{nunn2017theory}, and then it is mapped to noble-gas spins via weak spin-exchange collisions \cite{PRXQuantum.3.010305,PhysRevA.105.042606}. We consider the single-photon-based protocol \cite{RevModPhys.83.33} where single-photon sources and quantum memories are used for entanglement generation and swapping. We propose to use the same hot alkali atomic ensembles for single-photon sources. We quantify and analyze the entanglement generation efficiency and fidelity between two remote ensembles of noble-gas spins. Then, we show how entanglement swapping can be done to extend the entanglement to longer distances. Finally, we compute the repeater rates and overall fidelities and compare them to quantum repeaters with NV centers and optomechanics.

This paper is organized as follows. In Sec. \hyperref[sec:quantum system]{II}, we introduce the hybrid system of hot vapor and noble-gas spins in a ring cavity. The single-photon protocol is presented in Sec. \hyperref[sec:single]{III}. Sec. \hyperref[sec:rates]{IV} discusses the repeater rates and fidelities. Sec. \hyperref[sec:implementation]{V} gives more details on system implementation. We conclude and provide an outlook in Sec. \hyperref[sec:conclusion]{VI}.

\section{Hybrid atomic gas system} 
\label{sec:quantum system}

\begin{figure}
\centering
  \includegraphics[width=\linewidth]{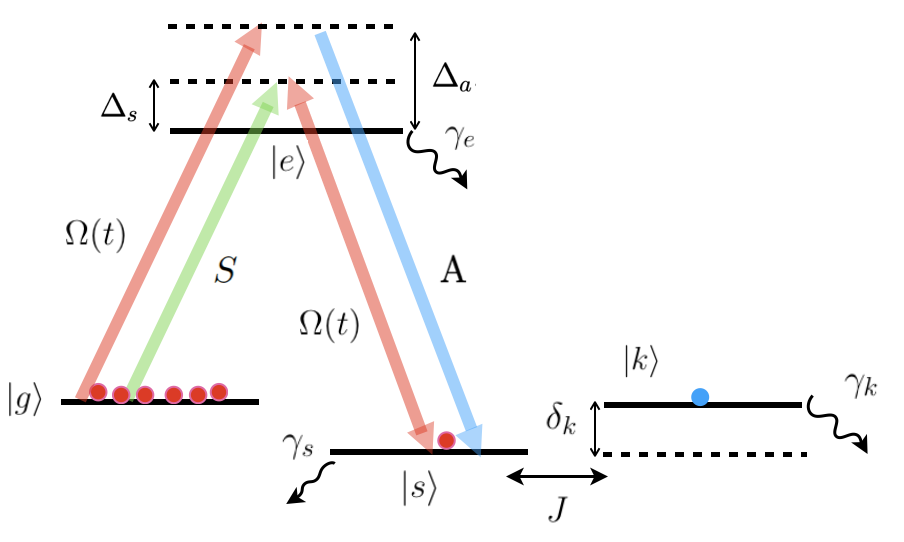}
  \caption{The level diagram of a hybrid system with hot vapor and noble gas. The $\ket{g}$-$\ket{e}$ transition is coupled by the input signal (Stokes field) with the strength proportional to $\sqrt{n_a}$ where $n_a$ is the number density of alkali atoms in the cell. The control field is coupled to the $\ket{e}$-$\ket{s}$ transition with a time-dependent Rabi frequency $\Omega(t)$. Both fields are detuned from $\ket{e}$ by $\Delta_s$, and the control field can also couple the $\ket{g}$-$\ket{e}$ transition with the detuning $\Delta_a$, which generates the anti-Stokes field A. The collective spin state for noble-gas atoms is denoted by $\ket{k}$. The $\ket{s}-\ket{k}$ transition is coupled to each other via the spin-exchange collision with a constant strength $J$ when these two states are in resonance with other, i.e. $\delta_k=0$. The collective excited state, alkali spin state, and noble-gas spin state decohere at the rates of $\gamma_e$, $\gamma_s$, and $\gamma_k$ respectively. Typically, $\gamma_k\ll\gamma_s$ as noble-gas spins have extremely low decoherence rates.}
  \label{Figure 2}
  \end{figure}

\begin{figure*}
\centering
  \includegraphics[width=\linewidth]{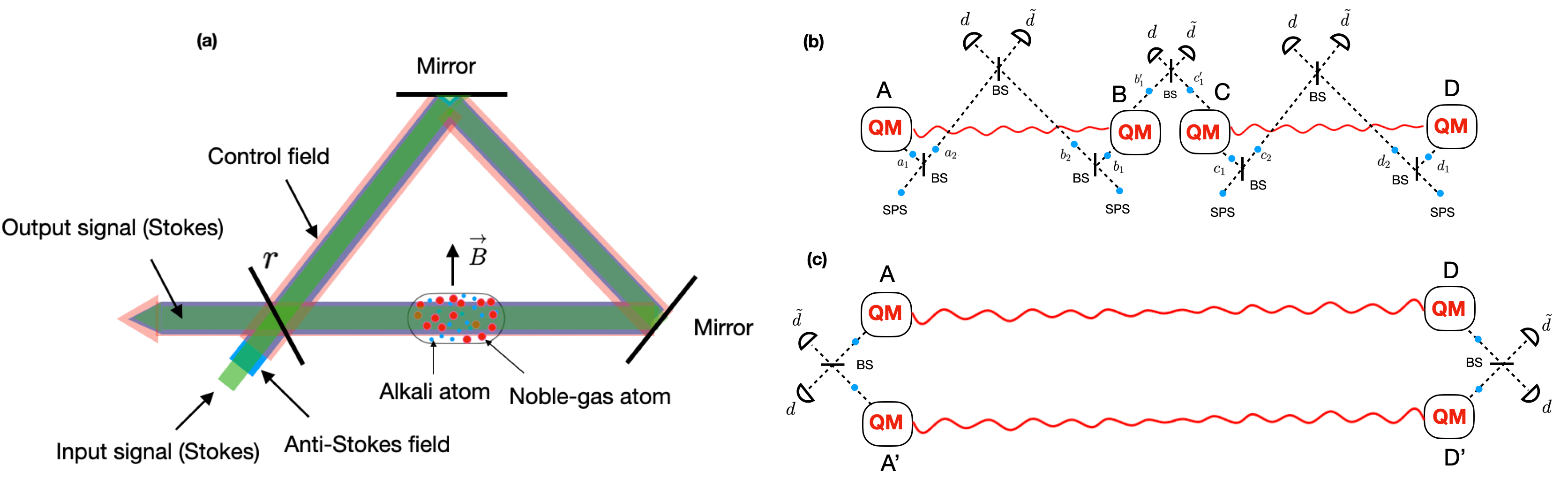}
  \caption{(a) Schematic of the setup of hybrid quantum memories. A cell that contains alkali atoms (red dots) and noble-gas atoms (blue dots) is placed inside a ring cavity where the green signal and red control field interact with alkali atoms directly. The FWM noise is the generated blue anti-Stokes field during the storage and retrieval processes, which can be largely suppressed via tuning the cavity. The interface between alkali atoms and noble-gas spins is based on spin-exchange collisions. (b) Schematic of a two-link repeater with single-photon sources \cite{PhysRevA.76.050301} as an example. There are two elementary links with four nodes where each one has a single-photon source which is a hot alkali atomic ensemble, a beam splitter, and a hot alkali-noble gases hybrid system serving as a quantum memory. The single-photon source emits a photon that either transmits through the beam splitter or gets reflected to enter the quantum memory. Two steps are required to establish the entanglement between nodes A and D. The first step is to generate the entanglement between A and B, C and D. The second step is to perform the entanglement swapping between B and C to distribute the entanglement to A and D. (c) Schematic of post-selection. This is the same as in the DLCZ protocol \cite{Duan2001}. The entanglement is established both in links A-D and A'-D', and the stored photons are retrieved to be detected, which allows us to rotate the measurement basis by adjusting the transmission coefficients and phases of the beam splitters \cite{RevModPhys.83.33}.}
  \label{Figure 3}
  \end{figure*}
  

As shown in Fig. \ref{Figure 3}(a), the hybrid atomic gas system is composed of a ring cavity and a cell of two hot atomic gases: alkali atoms and noble-gas atoms. This cell placed inside the cavity is driven by the control field (red) and the Stokes field (green). The ring cavity consists of two fully reflective mirrors and a mirror which serves as input-output coupler with amplitude reflectivity $r$. The coherent interaction between the noble-gas spins and alkali atoms is achieved by spin-exchange collisions \cite{PRXQuantum.3.010305}. As shown in Fig.\ref{Figure 2}, an ensemble of alkali atoms is modelled as a $\Lambda$-type system with a collective ground state $\ket{g}$, a collective spin state $\ket{s}$ and an excited state $\ket{e}$. Each noble-gas atom is modelled as a spin-1/2 system with up and down states $\ket{\Uparrow}$, $\ket{\Downarrow}$. Here, we denote the collective noble-gas spin state as $\ket{k}$. The input signal (Stokes field) $S$ couples the $\ket{g}-\ket{e}$ transition with the strength proportional to the density of alkali atoms $n_a$, and the control field couples to the $\ket{e}-\ket{s}$ transition with the Rabi frequency $\Omega(t)$. Both fields are detuned from $\ket{e}$ by $\Delta_s$. The control field can also couple the $\ket{g}$-$\ket{e}$ transition with the detuning of $\Delta_a$, which generates the anti-Stokes field A (FWM noise) because here all alkali atoms are prepared in one of the ground states that has higher energy \cite{nunn2017theory}. Due to the effect of spatial diffusion, there could be many spatial modes for the alkali and noble gases. However, in the light-dominated regime where the power broadening (proportional to the control field strength $|\Omega|^2$ and the optical depth $d$) in the alkali atoms due to the control beam dominates over diffusion in the alkali atoms, the collective spin mode of the alkali gas and the collective spin mode of the noble gas can be well approximated as single uniform modes by engineering the spatial profile of the control field \cite{PhysRevA.105.042606}. This condition is generally satisfied when both $|\Omega|^2$ and $d$ are large enough to be in the strong coupling regime, which is the case in this work as discussed in Sec. \ref{sec:implementation}.

After polarizing both the alkali and noble gases along the vertical axis using standard optical pumping and spin-exchange optical pumping (SEOP), the collective alkali spin state $\ket{s}$ can be coupled to the collective noble-gas spin state $\ket{k}$ via weak spin-exchange collision with the strength $J$. The coupling strength $J$ is given by $J=\zeta\sqrt{(2I+1)p_ap_bn_an_b/4}$ where $\zeta$ is the local average interaction strength of an alkali-noble atom pair in a single collision, and $p_a$ and $p_b$ are the polarization degrees of alkali and noble gases, and $n_a$ and $n_b$ are the densities of alkali and noble gases in the cell. $I$ is the nuclear spin of an alkali atom. Thus, $J$ is the effective interaction strength in multiple collisions with each collision averaging over all alkali-noble atom pairs in the ensembles \cite{PRXQuantum.3.010305}. $\delta_k$ is the detuning between these two states, which can be tuned by applying a magnetic field along the vertical axis. This detuning can be used to decouple these two species of atoms  \cite{PRXQuantum.3.010305,shaham2022}. $\gamma_e$, $\gamma_s$, and $\gamma_k$ are the decoherence rates for the collective excited state, spin state and noble-gas spin state respectively. Moreover, we have $\gamma_k\ll\gamma_s$ as noble-gas spins have an extremely low decoherence rate.

The Maxwell-Bloch equations of this hybrid system with the excited state $\ket{e}$ being adiabatically eliminated take the form  \cite{nunn2017theory,PhysRevA.105.042606}
\begin{equation}
\begin{aligned}
 (c\partial_z+\partial_t)S&=ic\sqrt{\frac{d\gamma_e}{L}}\frac{\Omega}{\Gamma_s}B-\kappa_sS,\\
(c\partial_z+\partial_t)A&=ic\sqrt{\frac{d\gamma_e}{L}}\frac{\Omega}{\Gamma_a}B^\dagger-\kappa_aA,\\
\partial_t B&=-i\sqrt{\frac{d\gamma_e}{L}}\frac{\Omega^*}{\Gamma_s}S+i\sqrt{\frac{d\gamma_e}{L}}\frac{\Omega}{\Gamma_a}A\\
&-(\frac{1}{\Gamma_s}+\frac{1}{\Gamma^*_a})|\Omega|^2B-\gamma_sB-iJK,\\
\partial_t K&=-(\gamma_k+i\delta_k)K-iJB,     
\end{aligned}
\label{MBeqs}
\end{equation}
where $S, A, B$, and $K$ are the annihilation operators for the signal field, anti-Stokes field, bosonic collective spin wave and collective noble-gas spin wave (we use the same notation as in \cite{PRXQuantum.3.010305}). $\Gamma_{s,a}=\gamma_e-i\Delta_{s,a}$ is the complex detuning of the signal and anti-Stokes fields. $d=g^2p_an_aL/\gamma_e$ is the optical depth where the alkali density $n_a$ is only nonzero in the cell, and $g$ is the average coupling strength between the Stokes/anti-Stokes fields and the alkali atoms, which is given by $g=\sqrt{\frac{1}{N_a}\sum^{N_a}_{i=1}{|g_i(r_i)|^2}}$ with $N_a$ being the number of the alkali atoms in the cell. This approximation is valid when the number of excitations is much smaller than $N_a$ \cite{PhysRevLett.105.140502}, which is the case here. $L$ is the length of a roundtrip in the cavity, and $c$ is the speed of light. The coordinate $z$ indicates the direction along the optical path inside the cavity. Moreover, the bosonic operators $B$ and $K$ take the form as $B=\sum^{}_{j\in \delta z}{\ket{g}_j\bra{s}/(\delta z\sqrt{p_an_a})}$ and $K=\sum^{}_{i\in\delta z}{\ket{\Downarrow}_i\bra{\Uparrow}}/(\delta z\sqrt{p_bn_b})$. As mentioned before, the spin-exchange coupling rate $J$ is proportional to the densities of the two atomic gases $n_a$ and $n_b$, i.e. $J\propto \sqrt{n_a n_b}$ \cite{PRXQuantum.3.010305}. Thus, by increasing the pressure, one can increase this interaction strength. $\kappa_s=cd\gamma_e/(L\Gamma_s)$ and $\kappa_a=cd\gamma_e/(L\Gamma^+_a)$ stand for the decay rates of the Stokes field and anti-Stokes field in the ring cavity where $\Gamma^+_a=\gamma_e-i(\Delta_a+\delta_s)$ with $\delta_s$ being the splitting between the states $\ket{g}$ and $\ket{s}$. Strictly speaking, Eq. (\ref{MBeqs}) should also have the Langevin noise operators. However, for both the signal and the anti-Stokes field, the noise is vacuum which is zero in normal ordering \cite{PhysRevA.76.033804}. Now, we have the boundary condition where the intra-cavity fields at $z=0$: $S_0$ and $A_0$ can be related to the input fields $S_{\text{in}}$, $A_{\text{in}}$ and the intra-cavity fields at $z=L$: $S_L$ and $A_L$ by the input-output coupler. Thus, we obtain the following relations \cite{nunn2017theory}:
\begin{equation}
\begin{aligned}
 S_0&=re^{ik_sL}S_L+t_rS_{\text{in}},\\
 A_0&=re^{ik_aL}A_L+t_rA_{\text{in}},
\end{aligned}    
\end{equation}
where $t_r=\sqrt{1-r^2}$ is the transmission coefficient of the coupler, and $k_s$ and $k_a$ are the wavevectors of the signal and the anti-Stokes fields respectively. Moreover, $S_L$ and $A_L$ can be directly related to $S_0$ and $A_0$ by Taylor expansion, which gives us:
\begin{equation}
\begin{aligned}
 S_L&\approx e^{-k_s\frac{L}{c}}(S_0+iL\sqrt{\frac{d\gamma_e}{L}}\frac{\Omega}{\Gamma_s}B_0-\frac{L}{c}\partial_tS_0),\\
 A_L&\approx e^{-k_a\frac{L}{c}}(A_0+iL\sqrt{\frac{d\gamma_e}{L}}\frac{\Omega}{\Gamma_a}B^\dagger_0-\frac{L}{c}\partial_tA_0),
\end{aligned}    
\end{equation}
where $B_0$ is the collective alkali spin operator for $z=0$. Combining these two sets of relations, one can obtain the following Maxwell-Bloch equations: 
\begin{equation}
\begin{aligned}
 \partial_ts&=-\tilde{\kappa}_s s+i\sqrt{\frac{d\gamma_e}{\tau}}\frac{\Omega}{\Gamma_s}b+e^{-i\phi_s}\frac{t_r}{\mu_s\sqrt{\tau}}S_{\text{in}},\\
\partial_ta&=-\tilde{\kappa}_a a+i\sqrt{\frac{d\gamma_e}{\tau}}\frac{\Omega}{\Gamma_a}a+e^{-i\phi_a}\frac{t_r}{\mu_a\sqrt{\tau}}A_{\text{in}},\\
\partial_t b&=-\gamma_s b+i\sqrt{\frac{d\gamma_e}{\tau}}(-\frac{\Omega^*}{\Gamma_s}s+\frac{\Omega}{\Gamma_a}a)\\
&-(\frac{1}{\Gamma_s}+\frac{1}{\Gamma^*_a})|\Omega|^2b-iJk,\\
\partial_t k&=-(\gamma_k+i\delta_k)k-iJb,     
\end{aligned}
\label{BMBeqs}
\end{equation}
where $\tau=L/c$ is the cavity roundtrip time. $s=\sqrt{\tau}S_0, a=\sqrt{\tau}A_0, b=\sqrt{L}B_0$, and $k=\sqrt{L}K_0$ are the intra-cavity amplitudes for the signal, anti-Stokes field, collective alkali spin state, and collective noble-gas spin state respectively. $\tilde{\kappa}_{s,a}$ is the resonant and anti-resonant decay rates for the signal and anti-Stokes field. They are given by \cite{nunn2017theory}:
\begin{equation}
\frac{1}{\tilde{\kappa}_{s,a}}=\tau\frac{\mu_{s,a}e^{i\phi_{s,a}}}{1-\mu_{s,a}e^{i\phi_{s,a}}},    
\end{equation}
where $\phi_{s,a}=k_{s,a}L-\Im{\kappa_{s,a}}\tau$ is the accumulated phases in the cavity roundtrip by the signal and anti-Stokes fields, and $\mu_{s,a}=re^{-\Re{\kappa_{s,a}}\tau}$ is the cavity roundtrip amplitude transmission for the fields.  


Eq. (\ref{BMBeqs}) can be solved in the bad-cavity regime where the signal/anti-Stokes field evolved at a rate much slower than the corresponding decay rate, i.e. $|\tilde{\kappa}_{s,a}|\gg|\sqrt{d\gamma_e/\tau}\Omega/\Gamma_{s,a}|$ \cite{PhysRevA.76.033804}. In this limit, we can set $\partial_ta\approx 0$ and $\partial_ts\approx 0$. Moreover, as we can  decouple the alkali and noble gases by applying a large magnetic field, we can break the storage into two steps: first consider the storage in the alkali atoms in the presence of the anti-Stokes field and then consider the transfer from the alkali to the noble gas. This sequential storage is optimal when the signal pulse duration $T$ satisfies $T\ll 1/\gamma_s$ \cite{PhysRevA.105.042606}, which is adopted in this work. We discuss how this sequential storage is achieved in detail and the optimal storage efficiency in Sec. \ref{subsec:EG}. The main noise present in the system is FWM, and in order to achieve the maximum suppression of this noise, we need to tune the ring cavity to be in resonance with the signal and to be in anti-resonance with the anti-Stokes field, which means $\phi_s=0$ and $\phi_a=\pi$. This is crucial in the first step of storage and retrieval, which is discussed in more detail in Sec. \ref{subsec:ES}. Other sources of noise in hot vapor systems include collision-induced fluorescence noise, the Doppler broadening, and inhomogeneous broadening for the $\ket{g}-\ket{e}$ transition. However, in this system, we ignore these effects as it has been demonstrated that fluorescence noise is negligible for the off-resonant scheme with a short pulse input \cite{Michelberger_2015}, and this is also true for the Doppler broadening and inhomogeneous broadening with the detuning $\Delta_s$ much larger than their bandwidth, which is discussed in Sec. \ref{sec:implementation}.

\section{The single-photon repeater}\label{sec:single}

Here, we focus on the single-photon-based protocol \cite{PhysRevA.76.050301} for entanglement generation and entanglement swapping where each node consists of a beam splitter (BS), a single-photon source (SPS) and a hybrid quantum system as depicted in Fig. \ref{Figure 3}(b), where we just show a two-link repeater as an example. As the noble-gas spins offer ultralong coherence time at room temperature, they will be used as the memory for storing the signal. There are two steps to establish the entanglement between two remote locations, and Fig. \ref{Figure 3}(a) shows how to achieve this between nodes A and D by cutting this distance into small pieces of equal length. In Fig. \ref{Figure 3}(b), it is cut into two equal pieces: A-B and C-D but it can be more general to have more links. In this example, we need to first establish the entanglement between A and B, and C and D, which is called entanglement generation, and then we perform the entanglement swapping between two local memories B and C to distribute the entanglement to A and D, i.e. only entangling A and D. Moreover, as a single excitation in the noble-gas spins is shared between A and D, it is difficult to perform measurements in other bases than the basis $\{\ket{k}$, $\ket{0}\}$. In order to relax this, we can introduce another entangled link $A'-D'$ where nodes $A'$ and $D'$ are in the same locations as $A$ and $D$ respectively \cite{PhysRevA.76.050301}, which is depicted in Fig. \ref{Figure 3}(c). In this way, we can use two beam splitters and two detectors in each location to read out the stored photons, which allows measurements in an arbitrary basis by choosing the transmission coefficients and phases. This step is known as post-selection. In this section, we show how entanglement generation, entanglement swapping and post-selection can be achieved in our hybrid system, and we also quantify the established entanglement generation fidelity and efficiency in the elementary link.

\subsection{Entanglement generation}\label{subsec:EG}
Before we characterize how the entanglement generation can be done, we would like to first talk about how signal storage can be achieved and give optimal storage efficiency. Our goal is to store the signal in the quantum memory as a collective excitation in the noble-gas spins, and this process can be divided into two steps: storing the signal in the collective spin excitation of alkali atoms and transferring this excitation to the collective excitation in noble-gas spins. This sequential storage is optimal when the signal pulse duration $T$ satisfies $T\ll 1/\gamma_s$ \cite{PhysRevA.105.042606}. In order to execute the first step, we make the detuning $\delta_k$ between $\ket{s}$ and $\ket{k}$ large enough such that $\delta_k\gg J$, and when this condition is satisfied, the states $\ket{s}$ and $\ket{k}$ are decoupled from each other \cite{PRXQuantum.3.010305}. Then, this process is simply described by the first three equations in Eq. (\ref{BMBeqs}) with $J=0$. Given that the maximum suppression of noise is achieved by tuning the ring cavity to be in resonance with the signal and to be in anti-resonance with the anti-Stokes field ($\phi_s=0$ and $\phi_a=\pi$), it has been shown that the optimal storage efficiency in the first step is $\eta_{1}=1-\sqrt{d}\gamma_e/(\sqrt{2}\Delta_s)$ in the strong coupling regime (more details can be in Sec. \ref{subsec:ES}) and the far-detuned regime ($\Delta_s\gg\gamma_e$) without mode mismatch in the cavity \cite{nunn2017theory}. This efficiency could be achieved when using lossless optical components. The requirements for all the related parameters can be realized experimentally, which are discussed in Sec. \ref{sec:implementation}. The second step is to transfer the signal stored in the alkali atoms to the noble-gas spins. Thus, we need to turn off the control field $\Omega(t)$ and tune $\ket{k}$ on resonance with $\ket{s}$ to make them interact, which can be done using an external magnetic field \cite{shaham2022}. The efficiency of this transfer is maximized when the transfer time is set to be $\pi/(2J)$ and it is in the strong coherent coupling regime, i.e. $J\gg\gamma_s\gg\gamma_k$ \cite{PhysRevA.105.042606}. Then, we obtain the optimal transfer efficiency $\eta_{2}=\text{exp}(-\frac{\pi(\gamma_s+\gamma_k)}{2J})$, which gives us the total storage efficiency:
\begin{equation}
\eta_s=\eta_1\eta_2=(1-\frac{\sqrt{d}\gamma_e}{\sqrt{2}\Delta_s})\text{exp}(-\frac{\pi(\gamma_s+\gamma_k)}{2J}). 
\label{eq:se}
\end{equation}

Now, we shall see how entanglement can be established in an elementary link. There are two links illustrated in Fig. \ref{Figure 3}(b), and here we focus on the first link for describing how the entanglement generation is achieved. In this link, for the left node, a single photon emitted from the source after a beam splitter can be described as $(\alpha a^{\dagger}_1+\beta a^{\dagger}_2)\ket{0}$ where $\alpha$, $\beta$ are reflection and transmission amplitudes of a beam splitter, and they satisfy the relation $|\alpha|^2+|\beta|^2=1$. The same is true for the right node where the state of a single photon after a beam splitter is $(\alpha b^{\dagger}_1+\beta b^{\dagger}_2)\ket{0}$. Thus, the joint state is given by:
\begin{equation}
[\alpha^2 a^\dagger_1b^\dagger_1+ \alpha\beta(a^\dagger_1b^\dagger_2+a^\dagger_2b^\dagger_1)+\beta^2a^\dagger_2b^\dagger_2]\ket{0}.   
\end{equation}
The first term in this state is the case where both single photons are reflected to be stored in quantum memories, ideally yielding no heralding in detectors. However, the detector dark counts could potentially lead to spurious clicks, thus causing infidelity in the desired entangled state. This probability is given by $\epsilon_0(1-\epsilon_0)\alpha^4$ where $\epsilon_0$ is the probability of having no dark counts in detectors. Here, we take it into account, but later on, we will see that its effect can be negligible if we choose the detector and detection window time properly. The second and third terms are the main contributions to single photon heralding where $a^\dagger_1$ and $b^\dagger_1$ are to be stored in quantum memories. We use noble-gas nuclear spins as quantum memories where the storage of a single photon is achieved in two steps as described above. As the finite storage efficiency, $\eta_s$ could create vacuum components, we take it into consideration in this work. The probability of having this contribution is given by $\epsilon_0\alpha^2\beta^2\eta_t\eta_c\eta_d\eta_s$ where $\eta_t, \eta_d, \eta_c$ are the transmission, detection, and frequency conversion efficiencies. The last term could also lead to the single-photon detection event when one of the two photons gets lost in the transmission, thus creating vacuum components as well. As discussed in Sec. \ref{subsec:ES}, although the hindsight from post-selection tells us that the vacuum components can be eliminated, which seems to have no effect on overall fidelity, it still could decrease the overall repeater rates. This probability is given by $\epsilon_0\beta^4\eta_t\eta_c\eta_d(1-\eta_t\eta_c)$. Moreover, we assume that the probability that the single-photon source emits a photon is $p_1$, which depends on the source we use. 

Here, we choose to use the same hot alkali gas as a single-photon source, which can be charged with a single excitation via FWM process used in the DLCZ protocol \cite{Duan2001}, and this atomic excitation can then be reverted to emit a single photon. A few experimental works have been reported for using hot rubidium atoms to generate bright and indistinguishable photons \cite{Ripka2018,Davidson_2021}. In this way, we do not need to perform frequency conversion to match with the alkali gas we use in the system, but the frequency conversion is needed for long-distance communication, i.e. for $a_2$ and $b_2$. Using atomic ensembles to generate single photons could lead to multi-photon errors thus degrading the repeater fidelities. This is discussed in detail in Sec. \ref{sec:rates}. We envision using the reverse-proton exchange (RPE) PPLN waveguide technique to convert a single photon emitted from the source to a telecom photon, which can operate at room temperature with a conversion efficiency of $23\%$ for the 863 nm signal \cite{PhysRevApplied.14.034035} but it is promising to apply it to the signal of different wavelengths. Moreover, by choosing the proper waveguide mode filter and fibre type, one can greatly improve this conversion efficiency to $60\%$ \cite{PhysRevApplied.14.034035}, and we use a higher value of $80\%$ in Sec. \ref{sec:rates} for rates calculations. Also, we assume that the relative phase in two optical fibres remains stable. Practically, this requirement can be achieved by actively stabilizing the lengths of fibre \cite{RevModPhys.83.33}, or through the use of self-compensating Sagnac-type configurations \cite{PhysRevA.77.052325}.

After taking all these effects into account, the entanglement generation fidelity and efficiency of the state created by detecting a single photon in one of the detectors are given by:  
\begin{align}
F_{\text{gen}}&=\frac{\alpha^2\beta^2\eta_t\eta_c\eta_d\eta_s}{\beta^2\eta_t\eta_c\eta_d+(1-\epsilon_0)\alpha^4-\beta^4\eta^2_t\eta^2_c\eta_d},\\
\eta_{\text{gen}}&=2p_1(\epsilon_0\beta^2\eta_t\eta_c\eta_d+\epsilon_0(1-\epsilon_0)\alpha^4-\epsilon_0\beta^4\eta^2_t\eta^2_c\eta_d),
\label{eq: figeff}
\end{align}
where $\epsilon_0=\text{exp}(-\lambda T_d)$ with $\lambda$ being the dark count rate, and $T_d$ is the detection window time which is set to be the time duration of the signal, that is $T_d=T$. $\eta_t$ is the function of the length of an elementary link $L_0$, which takes the following form: $\eta_t=\text{exp}(-L_0/2L_{\text{att}})$ with $L_{\text{att}}=22$ km being the attenuation length for telecom photons. The factor of 2 in the efficiency expression comes from the fact that the detectors are symmetric, and the heralding in either of them contributes to the efficiency. We envision using silicon single-photon avalanche diodes (Si SPADs) \cite{doi:10.1063/5.0034458,doi:10.1063/5.0041984} and frequency conversion to detect telecom photons. Si SPADs combined with a monolithic integrated circuit of active quenching and active reset (AQAR) can enable detection efficiency as high as $75\%$ with dark count rates below 100 Hz at 785 nm \cite{doi:10.1063/5.0034458}. This type of detector can operate at non-cryogenic temperatures which only require a thermoelectric cooler. The parameters are taken to be $\alpha^2=0.84$, $\beta^2=0.16$, $\eta_d=0.6$, $\eta_c=0.8$, $\eta_s=0.938$, $T_d\sim 12.5$ ns (the signal bandwidth is around 80 MHz, which is compatible with the hot vapor bandwidth as discussed in Sec. \ref{sec:implementation}). In this regime, the term $(1-\epsilon_0)\alpha^4$ is a few orders of magnitude smaller than $\beta^2\eta_t\eta_c\eta_d$ so Eq. (\ref{eq: figeff}) can be approximately written as $F_{\text{gen}}\approx\alpha^2\eta_s$, and $\eta_{\text{gen}}\approx 2p_1\beta^2\eta_t\eta_c\eta_d$. Moreover, we can now write the entangled state for each elementary link as
\begin{equation}
\alpha^2\eta_s\ket{\psi_{ab}}\bra{\psi_{ab}}+[\alpha^2(1-\eta_s)+\beta^2]\ket{0}\bra{0},
\label{eq:entangleds}
\end{equation}
where $\ket{\psi_{ab}}=\frac{1}{\sqrt{2}}(\ket{k_a}\ket{0_b}+\ket{0_a}\ket{k_b})$. The storage inefficiency $1-\eta_s$ increases the vacuum component proportion, and therefore it decreases the repeater rates. Here, we assume the storage efficiency to be $93.8\%$. Such a high storage efficiency is possible to achieve, which is discussed in Sec. \ref{sec:implementation}. The required input pulse is short as it satisfies the condition $T\ll1/\gamma_s$, which is also the requirement for the optimal signal storage in noble-gas spins using the sequential scheme \cite{PhysRevA.105.042606}. Moreover, when we have two elementary links, there is some waiting time for both links to establish entanglement, and as noble-gas spins offer ultralong coherence time, the decoherence that happened during the waiting time is ignored. 



\subsection{Entanglement swapping}\label{subsec:ES}

After we successfully establish the entanglement in two adjacent elementary links as shown in Fig. \ref{Figure 3}(b), we then need to perform entanglement swapping to propagate the entanglement between A and D. This can be done by recalling the single photon stored in either quantum memories B or C that are in the same location, and the heralding at one of the beam splitters informs us of the success in the swapping process, leading to the entangled state shared between A and D. At this level, it is well known that the swapping probability takes the following form \cite{RevModPhys.83.33}:
\begin{equation}
P_1=\frac{p_1F_{\text{gen}}\eta}{2}(2-p_1F_{\text{gen}}\eta),  
\label{eq:esef}
\end{equation}
where $\eta=\eta_d\eta_r$ is the product of the detection efficiency and the retrieval efficiency. Here, the retrieval process happens in two phases as well. First, we map the excitation in noble-gas spins to the excitation in hot vapor via the spin-exchange interaction by turning on the magnetic field for the amount time of $\pi/(2J)$ \cite{PRXQuantum.3.010305, PhysRevA.105.042606}. Second, we need to read out the signal from the collective spin state of the hot vapor. In this process, we need to turn on the control field $\Omega(t)$ and decouple the hot vapor from the noble gas by applying an external magnetic field to detune $\ket{s}$ from $\ket{k}$. The efficiency of retrieving the signal from hot vapor is the same as $\eta_1$, and it only holds under the condition that the decoherence of $\ket{s}$ is negligible during this process, which is true as the decoherence happens on the time scale much slower than that of memory interactions \cite{PhysRevA.105.042606, nunn2017theory}. Thus, the overall retrieval efficiency is $\eta_r=\eta_s$, which is given in Eq. (\ref{eq:se}). 

Now, putting all together, we can further simplify Eq. (\ref{eq:esef}) as $P_1=p_1\alpha^2\eta_{\text{tot}}(1-\frac{1}{2}p_1\alpha^2\eta_{\text{tot}})$, where $\eta_{\text{tot}}=\eta_s\eta_r\eta_d$. If we have more than two elementary links, the entanglement swapping is nested, which requires higher levels of swapping. This leads to a more general expression for the success probability of entanglement swapping at the $i$th level \cite{RevModPhys.83.33}:
\begin{equation}
P_i= \frac{p_1\alpha^2\eta_{\text{tot}}}{2}\frac{[2^i-(2^i-1)p_1\alpha^2\eta_{\text{tot}}]}{[2^{i-1}-(2^{i-1}-1)p_1\alpha^2\eta_{\text{tot}}]^2}.   \label{eq:estot} 
\end{equation}
After the entanglement swapping, a single excitation in noble-gas spins is shared between two remote locations (in Fig. \ref{Figure 3}(b), it is between A and D). As mentioned before, we need to perform post-selection by reading out the stored photons in each location, which allows us to generate an effective state $1/\sqrt{2}(\ket{k_Ak_{D'}}+\ket{k_{A'}k_D})$. Here, the dark counts are negligible because of the short detection time $T_d$ as mentioned in Sec. \ref{subsec:EG}. Then, the success probability of performing this projection is given by \cite{RevModPhys.83.33}:

\begin{equation}
P_{ps}=\frac{(p_1\alpha^2\eta_{\text{tot}})}{2}\frac{1}{[2^i-(2^i-1)p_1\alpha^2\eta_{\text{tot}}]^2}.    
\end{equation}
This post-selection step enables us to eliminate the vacuum components in Eq. (\ref{eq:entangleds}) as it is impossible to detect a single photon on each side if both links are vacuum. Hence, the overall fidelity is not affected by the vacuum components in Eq. (\ref{eq:entangleds}) but as mentioned they have a significant impact on repeater rates.

In the retrieval process, FWM noise can be strongly suppressed by choosing $\phi_s=0$ (on resonance) and $\phi_a=\pi$ (anti-resonance). In the strong coupling regime, this noise can be quantified by calculating the $g^{(2)}_{\text{re}}$ function of the retrieved signal, which is equal to $2|x|^2\zeta_1|\Gamma_s|^2/|\Gamma_a|^2$ \cite{nunn2017theory} when there is no mode mismatching, and the input signal contains one photon. $x$ is the FWM noise suppression factor, which is given by
\begin{equation}
x\approx\frac{1-\mu_s}{2\mu_s}=\frac{1-re^{-d(\frac{\gamma_e}{\Delta_s})^2}}{2 r e^{-d(\frac{\gamma_e}{\Delta_s})^2}}.
\label{noise}
\end{equation}
$\zeta_1\gg 1$ is the dimensionless coupling strength between both the signal and anti-Stokes field and the alkali gas, which is proportional to the control field strength $|\Omega|^2$ and the optical depth $d$ (More details are discussed in Sec. \ref{sec:implementation}). Then, the readout fidelity is given by 
\begin{equation}
F_{\text{re}}=\frac{1}{1+\text{SNR}^{-1}},    
\end{equation}
where $\text{SNR}^{-1}=g^{(2)}_{\text{re}}/2$ is the signal-to-noise ratio \cite{nunn2017theory}. Here, we ignore the infidelity that comes from the detector's dark counts as the detection window time is assumed to be around $12.5$ ns. Using the parameters discussed in Sec. \ref{sec:implementation}, it is possible to have a readout fidelity as high as $98.6\%$.

\begin{figure}
\centering
  $\hspace{-70mm}\mathbf{(a)}$\\
    \includegraphics[scale=0.4]{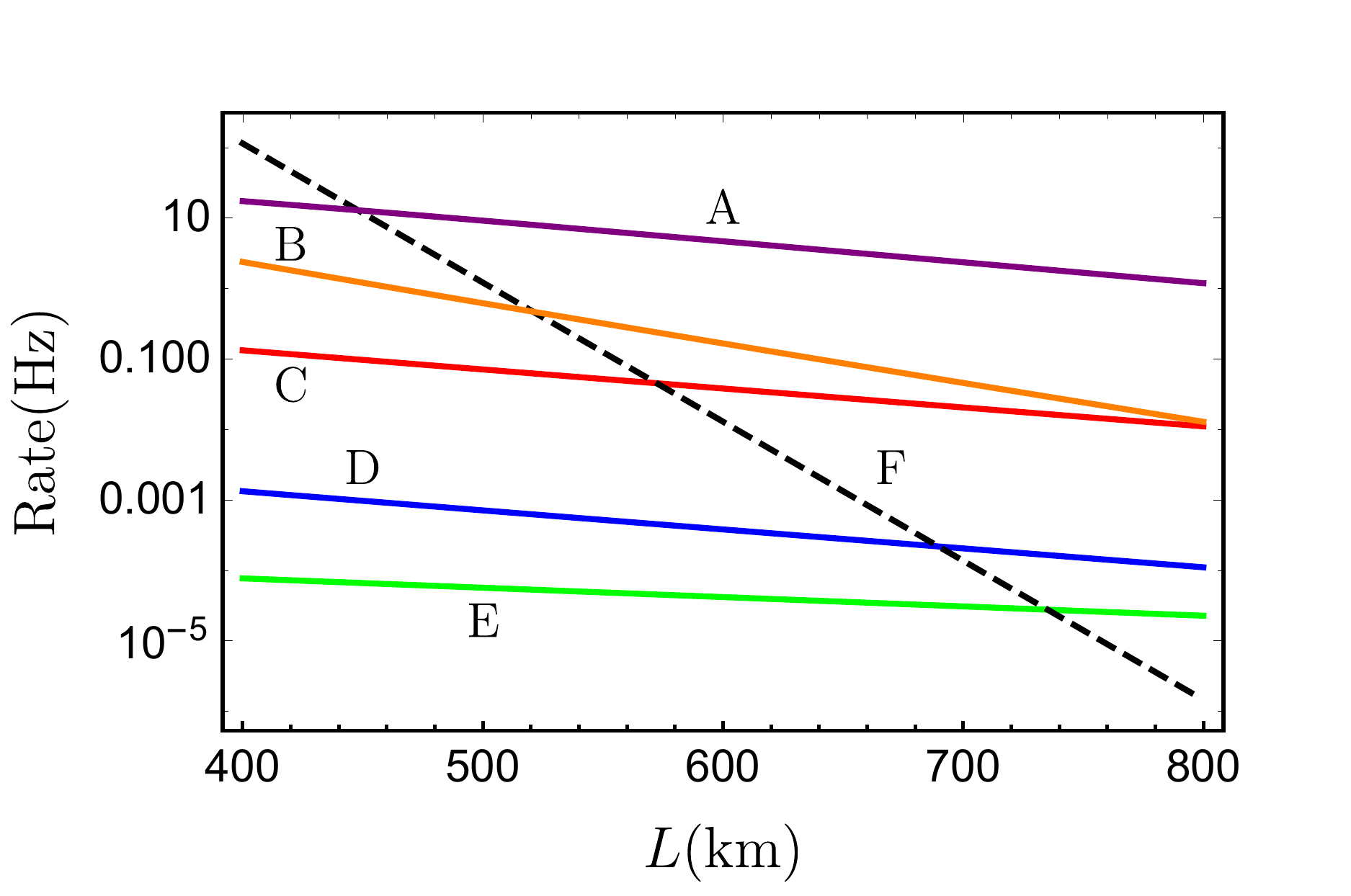}\\
    $\hspace{-70mm}\mathbf{(b)}$\\
    \includegraphics[scale=0.4]{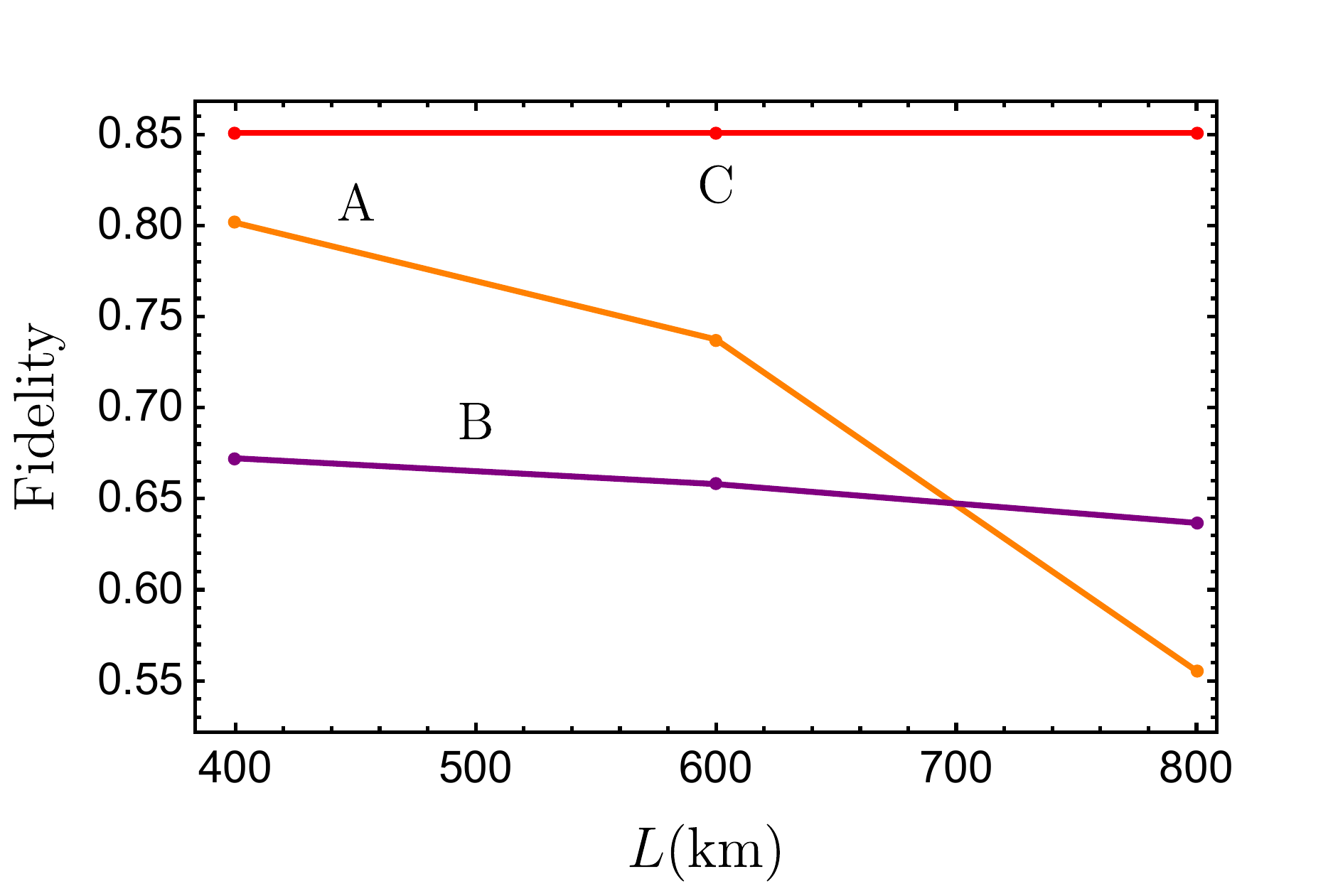}\\
  \caption{(a) Repeater rates as a function of total distance $L$ with $F_{\text{targ}}=0.9$ for hot hybrid gases-based scheme (referred to as scheme 1), and repeater rates for NV centers and optomechanics-based scheme (referred to as scheme 2) \cite{Ji2022proposalroom} with corresponding fidelities shown in (b). Here, we plot 4-link case (B), and 8-link case (A) for scheme 2, and we also plot 4-link case (D), multiplexed 4-link case (C), and 8-link case (E) for scheme 1. The direct transmission (F) is plotted with a single-photon source of $10$ GHz. For C, it is multiplexed by a factor of 100. In general, the rates of scheme 2 are much higher than the rates of scheme 1. All these repeaters outperform direct transmission. We assume $\eta_c=0.8$, $\eta_d=0.6$, $\eta_s=\eta_r=0.938$, $\alpha^2=0.84$, $\beta^2=0.16$, $t_{\text{trans}}=1.5$ ms, $t_{\text{ch}}=0.048$ ms and $1.03$ ms for D and E respectively. The emission probability for single-photon source $p_1$ is assumed to be 0.9 for all repeaters in both schemes. (b) Repeater fidelities as a function of total distance $L$ for schemes 1 and 2. A and B are 4-link and 8-link cases in scheme 2 \cite{Ji2022proposalroom}. C stands for a 100-multiplexed 4-link repeater in scheme 1 with $F_{\text{targ}}=90\%$ and $F_{\text{re}}=98.6\%$. As a multiplexed 8-link repeater in scheme 1 has fidelities very close to C, it is not shown here. Scheme 1 yields much higher fidelities than scheme 2, and they are independent of the total distance.}
  \label{Figure 4}
  \end{figure}

\section{Repeater rates and overall fidelities}\label{sec:rates}
In our system as the storage and retrieval time are mainly limited by how fast we can transfer the coherence from hot vapor to noble-gas spins via the spin-exchange collisions, and these times are given by $t_{\text{trans}}=\pi/2J$, which is around $1.5$ ms based on the parameters in Sec. \ref{sec:implementation}. This transfer time is on the same order as the two-way communication time $L_0/c$ for $L_0$ ranging from 50 km to 100 km with $c=2\times10^8$ m/s, which makes the total length of an 8-link repeater ranging 400 km to 800 km. Furthermore, the average charging time $t_{\text{ch}}$ in the ensemble also needs to be taken into account as it is comparable to $t_{\text{trans}}$ both in a four-link repeater and an eight-link repeater with the final target fidelity $F_{\text{targ}}=0.9$ as discussed later in this Section. Now, taking $\eta_{\text{gen}}$, $P_i$, and $P_{ps}$ into the standard entanglement distribution time for the single-photon protocol \cite{RevModPhys.83.33} plus the extra time spent for retrieving the signal and charging the ensemble, we obtain
\begin{equation}
T_{\text{tot}}= \frac{3^{n+1}}{2}(\frac{L_0}{c}+t_{\text{trans}}+t_{\text{ch}})\frac{\prod_{i=1}^{n}{(2^i-(2^i-1)p_1\alpha^2\eta_{\text{tot}})}}{\eta_t\eta_c\eta_dp^{n+3}_1\beta^2\alpha^{2n+4}\eta^{n+2}_{\text{tot}}}.
\label{eq:rrates}
\end{equation}
where $n$ indicates the number of nesting levels, and the number of links associated with it is $2^n$. Thus, the total length of a repeater is $L=2^nL_0$. In Fig. \ref{Figure 4}(a), we plot 4-link case (D), multiplexed 4-link case (C), and 8-link case (E) for hot hybrid gases-based scheme (referred to as scheme 1) with respect to total distance when $F_{\text{targ}}=0.9$ as discussed below. Also, we plot 4-link case (B), and 8-link case (A) for NV centers and optomechanics-based scheme (referred to as scheme 2) \cite{Ji2022proposalroom} for comparison. The direct transmission (F) is plotted with a source of $10$ GHz. For C, it is multiplexed by a factor of 100, which can be implemented spatially \cite{PhysRevLett.98.060502,RevModPhys.83.33} or spectrally \cite{KimiaeeAsadi2018quantumrepeaters} as discussed in Sec. \ref{sec:implementation}. All these repeaters outperform direct transmission at some point but in general, the rates of scheme 2 are much higher than the rates of scheme 1. The generally lower rates for scheme 1 are due to the fact that the single-photon protocol is nested as the entanglement swapping and post-selection are probabilistic as opposed to the non-nested scheme used in \cite{Ji2022proposalroom}, and the interface between alkali atoms and noble-gas spins is also quite slow, which further degrades the repeater rates. The other factors that limit the repeater rates in this proposal are detection efficiency $\eta_d$ and frequency conversion efficiency $\eta_c$, which could be improved to further enhance the rates. We expect an order of magnitude increase in rates when we increase $\eta_d$ from 0.6 to 0.9. The slow interface between hot vapor and noble-gas spins also plays a role in reducing the rates, but the room for improving the speed of this interface is limited as it is based on weak spin-exchange interactions \cite{PRXQuantum.3.010305}, which means $J$ cannot be too large. It is worth noticing that it is much easier to perform multiplexing in hot hybrid gases-based repeaters than NV centers and optomechanics-based repeaters because the latter requires much more complex setups than the former \cite{Ji2022proposalroom}. Moreover, there is a trade-off between the target fidelity $F_{\text{targ}}$ and repeater rates as $F_{\text{targ}}$ determines $t_{\text{ch}}$. However, the improvement in rates is not significant when we set a lower target fidelity.

The infidelities in our repeaters mainly come from multiphoton emissions of the single-photon source and FWM noise in the entanglement swapping and post-selection. The effect of FWM noise in the signal readout has been estimated in Sec. \ref{subsec:ES} based on the parameters discussed in Sec. \ref{sec:implementation}, which gives us a high readout fidelity of $98.6\%$. In addition, since we use noble-gas spins as quantum memories, which have hours-long coherence time, thus the infidelity induced by spin decoherence can be negligible as long as the repeater rate is above the order of $10^{-3}$ Hz. This is true for multiplexed repeaters and non-multiplexed 4-link repeaters for $L$ up to 800 km as shown in Fig. \ref{Figure 4}(a). Now, the overall fidelity is given by
\begin{equation}
F_{\text{tot}}=F_{\text{targ}}\times (F_{\text{re}})^{n+2},    
\end{equation}
where $n+2$ is the number of performing readouts. $F_{\text{targ}}$ is a target fidelity of repeaters, which we choose to be $90\%$ for all repeaters in scheme 1 with different nesting levels. This fidelity is determined by errors due to multiphoton emission in the ensemble-based single-photon source \cite{RevModPhys.83.33}. The probability of having a two-photon contribution is given by $p_2=2p(1-\eta_{\text{st}})p_1$ where $p$ is the probability of emitting the Stokes photon when charging the ensemble, and $\eta_{\text{st}}$ is the efficiency of detecting a Stokes photon, assumed to be 0.75 using a silicon single-photon detector \cite{doi:10.1063/5.0034458,doi:10.1063/5.0041984}. In order to make $p_2$ small enough to have $F_{\text{targ}}=0.9$, we need to make $p$ sufficiently small. It can be shown that when we have a four-link repeater, the maximum value that $p_2$ can take is $0.00093$ \cite{RevModPhys.83.33}, which leads to $p=0.0021$. This emission probability results in a charging time given by $t_{\text{ch}}=1/(Rp)=0.048$ ms with the repetition rate $R=10$ MHz. If we have an eight-link repeater with $F_{\text{targ}}=0.9$, we obtain $p=9.73\times 10^{-5}$, which leads to $t_{\text{ch}}=1.03$ ms. Assuming the readout fidelity for both swapping and post-selection is $98.6\%$, the overall fidelities of a 4-link and 8-link repeaters in scheme 1 are estimated to be $85.1\%$ and $83.87\%$. In Fig. \ref{Figure 4}(b), we plot the overall fidelities as a function of total distance $L$ for scheme 1 and scheme 2. A and B are 4-link and 8-link repeaters in scheme 2 which decrease as total distance increases due to thermal noise present in the system which are treated as dark counts \cite{Ji2022proposalroom}. C is a 4-link repeater in scheme 1, which is independent of the total distance. In general, scheme 1 yields fidelities that are significantly higher than the fidelities in scheme 2, which is mainly due to the fact that the accumulated infidelities induced by vacuum components are eliminated in the end by post-selection. Overall, these two schemes have their own advantages and disadvantages. Scheme 1 is much slower than scheme 2 but has much higher fidelities, and scheme 1 requires much less complex setups than scheme 2 which also facilitates multiplexing. Moreover, it is possible to boost the fidelities using entanglement purification \cite{PhysRevLett.104.180504}, but this comes at the cost of further reducing the rates. A quantitative discussion of repeaters including purification goes beyond the scope of the present work.


\section{Implementation}\label{sec:implementation}
Here, we consider \ce{^{39} K} atoms as the hot vapor and $\ce{^{3} He}$ atoms as the noble-gas spins in our system, where the optical depth $d$ of the hot vapor is assumed to be 100, which can be achieved by choosing the length of the cell given that the temperature of hot vapors is fixed. The linewidth of the excited state $2\gamma_e$ is taken to be $27$ GHz for broadened $\text{D}_1$ line due to collisions with buffer gas, which is much smaller than the assumed detuning $\Delta_s=2700$ GHz so it makes the system in the far-off resonant regime \cite{nunn2017theory}. Moreover, such a large detuning $\Delta_s$ also makes the Doppler broadening negligible which is around 1 GHz at $230^\circ$ C. So far, the experimentally achieved value of $J$ is around $78$ Hz \cite{shaham2022} but if we further increase the pressure to increase the gas densities, it is possible to have $J=1000$ Hz \cite{PRXQuantum.3.010305}. In this condition, for $\gamma_s$ and $\gamma_k$, they are estimated to be $17.5$ Hz and $10^{-4}$ Hz respectively, which is dominated by intra-gas and inter-gas collisional spin-rotation couplings \cite{PRXQuantum.3.010305,katz2021coupling,shaham2022}. Thus, this yields an optimal storage efficiency of around $93.8\%$. Moreover, $\Delta_a=\Delta_s+\delta_s$ where $\delta_s$ is the splitting between the states $\ket{g}$ and $\ket{s}$, which is around 0.46 GHz in $\ce{^{39} K}$ vapor. In the strong coupling regime ($\zeta_1\gg 1$), the noise suppression factor $x$ is given in Eq. (\ref{noise}), and when the storage and retrieval efficiencies are optimized, the reflectivity $r$ is given by $r=(1-\sqrt{1-\alpha^2_s})/\alpha_s$ with $\alpha_s\approx \exp\{-d(\gamma_e/\Delta_s)^2\}$, which is estimated to be $93.2\%$. Thus, we obtain the signal readout fidelity $F_{\text{re}}\sim 98.6\%$. The cavity linewidth $\kappa_c$ is linked to $r$ and the hyperfine splitting $\delta_s$ as $\kappa_c=8\delta_s(1-r)/r$, which is estimated to be $0.27$ GHz. Moreover, in the bad cavity regime, the bandwidth $\delta_B$ of this hybrid quantum memory is upper bounded by the cavity linewidth as $0.3\kappa_c$ \cite{nunn2017theory}, which gives $\delta_B\sim 80$ MHz. The size of the ring cavity is given by the length of roundtrip $L=\pi c/(2\delta_s)=160$ mm. As for the time-bandwidth product, this hybrid quantum memory yields an unprecedented value of $8\times 10^{11}$ which is mainly attributed to the hours-long storage time in the noble gas and the large bandwidth of the hot vapor. The multiplexing can be implemented either spatially or spectrally. For spatial multiplexing, we envision having many hybrid memories in each node \cite{PhysRevLett.98.060502}, which is possible thanks to the mm-scale system size. The spectral multiplexing also requires many hybrid memories in one node but the emitted photons need to be converted to different frequencies fed into a common channel \cite{KimiaeeAsadi2018quantumrepeaters,Glorieux:12}. This can be accomplished using frequency translation which can be noise-free using waveguide electro-optic modulators \cite{PhysRevLett.119.083601}. The feeding to a common channel can be achieved by a tunable ring resonator filter that enables MHz-level resonance linewidths \cite{Yang2018}.

\section{conclusions and outlook}\label{sec:conclusion}
We presented a quantum network architecture based on hot hybrid alkali-noble gases that can operate without cryogenics. We showed that under realistic conditions, high-fidelity entanglement can be distributed over long distances thanks to the ultra-long coherence time of noble-gas spins. We showed that the rates of our proposed quantum repeaters can outperform direct transmission, and with realistic multiplexing, the rates can be greatly enhanced, close to the corresponding rates of NV centers and optomechanics-based repeaters. Furthermore, our system requires no extra components other than a ring cavity. This significantly reduces the complexity of the system while offering a great potential to be scalable, which is more advantageous over the room-temperature repeaters with NV centers and optomechanics. We hope that this work could further stimulate the development of high-efficiency silicon single-photon detectors and even room-temperature detectors that offer both high detection efficiencies and low dark count rates for telecom photons.

We here have focused on hot atomic gas-based quantum repeaters on the ground, but this compact hybrid quantum system also offers a good potential for being used as memory in space \cite{gundougan2021topical}, which could unlock the possibility of establishing a truly global quantum network \cite{Gundogan2021,Liorni_2021, PhysRevA.91.052325} that goes beyond the limit of terrestrial quantum repeaters, and such a global quantum network could enable ultra-long distance quantum teleportation, quantum entanglement and applications in fundamental physics tests \cite{gundougan2021topical}.

\section*{Acknowledgements}
\label{sec:acknowledgements}
This work was supported by the Natural Sciences and Engineering Research Council of Canada (NSERC) through its Discovery Grant (DG), CREATE, and Strategic Project Grant (SPG) programs, and by the National Research Council (NRC) of Canada through its High-Throughput Secure Networks (HTSN) challenge program, and by Alberta Innovates Technology Futures (AITF) Graduate Student Scholarship (GSS) program. In the spirit of respect and reconciliation, J.-W.J., F.K.A., and C.S. acknowledge that the University of Calgary is located on the traditional territories of the people of the Treaty 7 region in Southern Alberta, which includes the Blackfoot Confederacy (comprising the Siksika, Piikani, and Kainai First Nations), as well as the Tsuut’ina First Nation, and the Stoney Nakoda (including the Chiniki, Bearspaw, and Wesley First Nations). The City of Calgary is also home to the Métis Nation of Alberta (Region 3). K.H. acknowledges that the NRC headquarters is located on the traditional unceded territory of the Algonquin Anishinaabe and Mohawk people.

\bibliography{mybib}{}

\end{document}